\documentstyle[aps,prl]{revtex}

\begin{document}
\draft
\title{
Roundoff-induced Coalescence of Chaotic trajectories }%
%
\author{Lech Longa\cite{AAA},  Evaldo M. F. Curado$^{,\ddagger}$ 
and  Fernando A. Oliveira }
%
\address{ International Centre of Condensed Matter 
         Physics, Universidade 
  de  Bras\'{\i}lia,   70919 - 970 Bras\'{\i}lia, DF, Brazil.\\
  $^{\ddagger}$Centro Brasileiro de Pesquisas F\'{\i}sicas, 22290-180,
  Rio de Janeiro, RJ, Brazil 
  }
\date{\today}
\maketitle
\begin{abstract}
Numerical experiments recently discussed 
in the literature show that 
identical nonlinear chaotic systems linked by 
a common noise term (or signal) may synchronize after a finite time. 
We study the process of synchronization as function of precision
of calculations.  Two generic behaviors
of the average   coalescence time are identified: 
exponential or linear. In both cases no synchronization  
occurs if iterations are done  with {\em infinite} precision.
\end{abstract}
\narrowtext

It is well known that  deterministic systems 
may  exhibit a chaotic behavior \cite{review}.
Typically, two chaotic  orbits initially 
displaced only slightly from each other
deviate exponentially with  time approaching 
separation   on the order of the strange 
attractor size. 
In practice it   means that small errors make 
longtime evolution of nonlinear, dynamical systems unpredictable
\cite{review}. Thus, identical chaotic systems
are not expected to synchronize. 

Recently, in a series of papers 
\cite{pecora,pikovsky0,maritan,pikovsky}, it has been 
demonstrated  that synchronization could  nevertheless be achieved 
if the systems are linked by a {\em common} signal or  noise term 
of an appropriate strength.  That is,  if we take  a chaotic map
and start numerical evolution of two arbitrarily chosen initial points
subject to the same sequence of  noise, the resulting trajectories will 
collapse after a finite number of iterations. 

Two cases could be distinguished. The first one,  discussed by 
Pecora {\it et al.} \cite{pecora} and Fahy
{\it et al.} \cite{pecora} (PF),  defines the collapse  
as a process where the average distance between the
trajectories converges exponentially to zero. 
According to PF  such 
synchronization takes place  if the largest Lyapunov 
exponent becomes negative. 

A different scenario was found
by Maritan and Banavar (MB) \cite{maritan}.
They claimed that the coalescence  also may  occur abruptly, 
without any specific time dependence of the average 
distance between the trajectories.
As an  example they studied  
the  well-known one-- dimensional logistic map 
\cite{review,feigenbaum,ott,evaldo,fernando} in   
the presence of additive noise 
\begin{equation} \label{logistic}
       x' = 4\,x\,(1-x)\,\, + \,\, W\,\eta
\end{equation}
where $ 0 \le x \le 1 $;  $\eta$ is a
random number chosen uniformly from the interval $-1$ to $+1$ and $W>0$ 
is the strength of the noise. The values of $\eta$ violating the bounds
$0 \le x' \le 1$ were discarded and a new $\eta$ was chosen.  
They found that for  $W > W_{c}$ ($W_{c} \approx 0.55$) 
and after about $10^{6}$ iterations  the
trajectories of almost any two points became identical 
and  follow a single random trajectory.  
As conclusion they claimed that in the limit of infinite time and finite
precision all trajectories should collapse for $W > W_{c}$.
Thus at $W=W_{c}$ one would get a kind of  phase transition.
Several  data were exhibited  to confirm
the collapse and the existence of the transition.

Recently Pikovski \cite{pikovsky} 
has suggested that the  synchronization observed by MB
is a numerical effect of finite precision of calculations.
However no systematic analysis has  been given.
It is therefore of interest to investigate the effect of 
computer roundoff on the coalescence, which we would 
like to propose in this letter.   We show that  the average 
coalescence time (or simply the relaxation time), $T$,  
for  the two orbits of the model (1)
follows  the exponential law:  $T = \mbox{e}^{ {A(W)}  
+ {B(W)} N}$ with  $A$ and $B$  being functions of 
the noise  strength;  $N$ is the precision of the calculations. 
This law, which is  different from the one 
suggested in \cite{pikovsky}, proves that statistically the coalescence   
{\em never} occurs for a map or flow of infinite precision.

We also show that {\em no nontrivial} critical value $W_{c}$  exists.
Even for  $W < 0.5$ the collapse does occur.  If
we  consider finite precision iterations 
and  $W \ge W_{min} = 0.5\mbox{\small x} 10^{-N}$
then the two different trajectories should always collapse. 
As conclusion one can say that the limit discussed in the paper of MB
(first, time going to infinity, then the precision going to 
infinity) is trivially satisfied in the sense that (almost) all 
two points coalesce in this limit.  
Finally, a synchronization by common signals
observed in the chaotic systems  with negative Lyapunov 
exponents \cite{pecora} yields the relaxation time  
linear  with $N$.

Our analysis starts  by considering 
the map  studied by MB, Eq.(\ref{logistic}).  
But the  approach  differs from 
that of MB  in the way  the map  is iterated. 
Namely, we perform  the iterations with {\em controlled absolute accuracy}  
by fixing the number of digits after decimal point. More specifically, 
for all numbers from  interval $[0,1]$ with decimal representation
$ x = 0 \mbox{{\Large\bf{}.}} a_{1}a_{2} \cdots a_{N}$ 
the value of $N$ is being fixed after each iteration.
Here the symbol ``{\Large\bf{}.}'' represents  decimal point and 
$a_{i} = 0,1,...,9$  ($i=1,...,N$).
For  $x=1$ we get automatically $x' = 0$.

Finite accuracy  calculations are done with  
programs  running  under control of the  {\sc Maple} package. 
To use  the program  the precision  $N$ has to be fixed.
We assume  $N$ to vary from 1 to 16.  Next, the noise strength $W$ 
must be given and the two 
initial numbers for the trajectories  chosen randomly 
from the interval $[0,1]$.
After these preparations are done we start two independent  
iterations according to the prescription (\ref{logistic}).
The process continues until both trajectories 
became numerically identical.
The corresponding number of iterations is called  coalescence time 
and denoted  $t$. The whole procedure (with fixed $N$ and $W$) is repeated 
100 times to get  $T$. 

Results of the simulations are shown as triangles in 
Fig.(1),  where  $\ln (T)$  varies with   $N$  for $W=0.7$. 
Circles represent similar simulations but with $N$ denoting {\em relative
precision i.e.}  the number of
digits in the mantissa as in usual floating point arithmetic.
Note that for the map given by  Eq.(1), 
the absolute-- and the relative precision
calculations yield statistically the same results.
 
Additionally, two histograms representing  
the probability distribution function $P(t)$ that the coalescence 
time is $t$, are shown as insets of Fig.(1). They are constructed  
for $N=8$ and for two different 
samples composed of 100 and 10000 points, respectively.
Though  $t$ has   substantial statistical fluctuations, the difference
between   $\ln (T)$ 
calculated for both samples differs by less than 0.9\%.
Similar result is found for $N=15$ with the averages  taken
over 100 and 1000 points.  Hence
100 realizations of the process  provide a useful estimate of $\ln(T)$.

We shall observe that  $T$ increases 
exponentially  with $N$.  This has been shown in Fig.(1)
by plotting a line  $\ln(T) \simeq 2.297 + 0.765\, N  $ ($16 \ge N \ge 3$)
of the best least - square fit, which on the average is good to 
within better than 2.5\%{}. Interestingly, similar linear dependence of  
$\ln (T)$ on N is predicted for other values of $W$. 
 This indicates that (statistically) 
two trajectories  {\em never} collapse in the limit of infinite 
precision.

At this point we would like  to turn to  the 
results of MB. They found coalescence of the trajectories for $W=0.6$ 
but  not for $W=0.3$, in both cases the number of iterations being 
$5\,\mbox{\small{}x}\,10^{5}$.  Clearly,  for $W=0.3$
 the number of iterations they performed 
was much too small to see coalescence.
We found that for double precision experiments and 
for $W=0.5$ -- $0.6$, time  $T$ is  {\em orders 
of magnitude smaller} than that for $W=0.1$ or $W=0.3$. 
More specifically, for $W=0.1$
the collapse is expected after about $10^{14}$ iterations while 
for $W=0.3$ after $10^{12}$ iterations.  This effect is illustrated 
in Fig.(2), where  $\ln (T)$ is plotted as a function of $W$ for 
different values of $N$. As previously averages are taken over 100 
$t$'s. Again the influence  of scatter of the data on  $\ln (T)$ is tested
for $N=3$ and $N=7$ where additionally the averages are 
taken over 1000 experiments. In both cases the results  are  
practically independent of the sample size.
For $N=3$ this is shown in Fig.(2) where triangles represent averages over 
100 experiments and circles over 1000.
It is  seen that $T$ is  high for small values of $W$ and decreases
with increasing $W$. A minimum of $T$ is  found
around $W=0.6$. Then $T$ increases  again to finally saturate 
for $W \ge 1$.  This behavior  is observed for   all
values of N  studied and its  origin  will be explained later in
the paper. 
 
Now one can easily understand the results of MB, presented 
in their Fig.(2).  A maximum of the probability that the two 
trajectories collapse (denoted as  $\lambda$ in \cite{maritan}),  
corresponds to our minimum of $T$
(see Fig.(2)).  Likewise, the decrease of $\lambda$ 
for  $W >0.6 $ is connected with our  increase of $T$.
The difference between our results and theirs 
is that we do not observe any transition around $W=0.5$.  
Generally we find that collapse takes place for all values 
of $W$ such that $W \ge W_{min}$.  This  has been checked 
for $N \le 8$.  For higher values of $N$ time $t$ becomes 
too long and  calculations are practically
impossible.

In order to understand  why $W$ must be nonzero  to get collapse 
please note that any chaotic map looses its chaotic character
when  iterated with a finite accuracy \cite{grebogi}. 
Take,  for example,  the logistic map (\ref{logistic})
and fix accuracy to $N=3$. The domain  of the  map 
consists in this case  $10^{3} + 1$  states (boxes) equal 
($0,0.001, 0.002, \cdots, 0.999,1$). For $W=0$  these numbers 
evolve either to one of the fixed points: $x = 0$ or $x=0.750$ or to 
a cycle of length 13: $\{ 0.109, $ $ 0.190, $ $ 0.204, $ $0.328, $ 
$ 0.388, $ $ 0.416, $ $
0.616, $ $0.650,$ $ 0.882, $ $0.910, $ $ 0.946, $
$ 0.950, $ $0.972 \}$.  Thus, in the absence of  noise 
the  trajectories belong to periodic orbits or fixed points and 
only accidentally  they can  collapse. 
For $W \ne 0$  the structure of the cycles 
is destroyed.  Even the smallest numerical value of the noise 
($W = W_{min}$), which either does not modify 
the iterated states  or modifies them 
by $\pm 0.001$  (each realization being of the same probability 
equal 1/3), is sufficient to have nonzero probability that  
all initial states are again accessible from any other state after  
a finite number of iterations.  Interestingly,  $W_{min}$ does not  depend  
on detailed structure of periodic orbits   indicating 
that any sort of noise that restores ergodicity would
yield  collapse of the trajectories. The same statements hold for 
arbitrary $N$. Instead, if we consider  infinite accuracy the probability 
that two numbers coming out of iterations  are different is one. 
As we are in   chaotic region, where dynamics is strongly sensitive
to the initial conditions, these two points will move away 
from each other in the forward iterations.

Another new feature of the  logistic map found in our simulations and 
displayed in Fig.(2) is   behavior of the
curves $\ln{}(T)$ as function of $N$. Namely they {\em all} scale 
linearly with $N$, {\em i.e.} $  \ln (T) = A(W) \, + \, B(W)\, N $, 
with $W$-dependent coefficients. 
Using  the data shown in Fig.(2)  we estimated $A(W)$  and $B(W)$
by  performing linear least square  fit  for each $W$ separately. 
The obtained set of points  is  given in Fig.(3). 
 These data show that for $W \approx  1 $ the time $T$ 
 behaves as   $10^{N/2}$ and for $W=0.6$ as 
 $10^{N/3}$, which differs from the $10^{2N}$--law
 suggested in \cite{pikovsky}.

Finally, the Fig.(3) allows one to estimate $T$ for $W=0.3$
 and for $N=15$ as being of order of $10^{12}$. This is the reason
 why in experiments of MB, which dealt with $10^{6}$ iterations,
 the collapse for $W=0.3$ statistically has not been observed. 

A further insight into the coalescence  process  and its connection 
to the accuracy  could  be achieved    by  
converting the logistic map (1) into the 2y modulo 1 map  \cite{ott}
\begin{equation} \label{shift}
 y' = 2y   \,\,\,\,\,\,\,  modulo\,\, 1  + g(W,\eta, \, ...).    
\end{equation}
This is done in practise by a change of variables: 
$x = \frac{1}{2}\big[ $ $ 1- cos(2\pi y) \big]$. 
Note that the transformed noise, denoted by $g(W, \eta, \, ...)$, depends 
at time $n$  on the system states  prior to n.   
This dependence, represented  by dots in Eq.(\ref{shift}),
appears quite nontrivial and will not be of our concern  here. 
Instead we will discuss a simple case where precise form of 
$g(W,\eta,\, ...)$ is irrelevant.  Again the values of $g$ 
violating the bonds  $0\le y' \le 1$ are discarded.

Suppose that  we study the binary version
of the  map (\ref{shift}) in the interval [0,1). That is we 
iterate the map by
representing 
each number (including the noise term) as a  binary decimal
 ${0} \mbox{{\Large\bf{}.}} a_{1}a_{2}a_{3} \cdots a_{N}\cdots \equiv
 \sum_{i=1}^{\infty}2^{-i}a_{i}$, where each of the digits $a_{i}$ is 
 either $0$ or $1$. With this simple change of base 
 the iterations of the map (\ref{shift}) could be viewed
as moving the decimal point ``{\Large\bf{}.}'' one position to the right
({\em Bernoulli shift}).  

 A binary number of accuracy $N$ could now be 
 introduced  as the one for which $a_{j} = 0, \,\,\,\, \forall  j > N$. 
Consider  two arbitrary  numbers $x$ and $y$ of binary accuracy $N$
and assume that  their  binary representations  
are of the form
 $0 \mbox{{\Large\bf{}.}} x_{1}x_{2}x_{3} \cdots x_{N} $
 and 
 $0 \mbox{{\Large\bf{}.}} y_{1}y_{2}y_{3} \cdots y_{N} $,
respectively. The first iteration moves the points one step to the right 
yielding 
 $x' = 0 \mbox{{\Large\bf{}.}} x_{2}x_{3} \cdots x_{N}0 $ and 
 $y' = 0 \mbox{{\Large\bf{}.}} y_{2}y_{3} \cdots y_{N}0 $.
 As the last digit (zero) is now the
same for both numbers x' and y' the noise converts them into two
numbers with the same digit (the $N$-th digit of the noise) 
at position $N$ after the  point.
Successive iterations in the presence of noise 
increase the number of identical digits by one making 
the coalescence time  $t \lesssim N$ and, consequently, $T$
must be a linear function    of $N$.  
Similar linear behaviour of $T$ with $N$
is predicted for the PF models  \cite{pecora}. 
This follows directly from the fact that
the average distance, $d$, between the PF trajectories at the 
time $T$ is $d = A_{0} 10^{-T/\lambda}$, where $A_{0}$ and 
$\lambda$ are model dependent parameters.
Comparing $d$  with $10^{-N}$  yields $T \sim \lambda N$.

Let us now come back to the logistic map and comment  
two issues: {\em (a)} why the minimal $T$ 
is found for $W \gtrsim 0.5$ and {\em (b)} how  $T$
could be connected with the properties of the map and of the noise. 
 An important point to note  (see also \cite{maritan}) is that,
in general, chaotic maps consist of iterations that are 
composed of two parts,  
the one which stretches the  distance between the points 
and the other  where the distance is  enlarged. Consider the distance
$d_{l}$ between images of $x$ and $y$ 
\begin{equation}
  d_{l}: {\cal{S}} \ni (x,y) \rightarrow 4|x-y|\,\, | 1 - (x+y) |,
\end{equation}
where ${\cal{S}} = [0,1]\times[0,1]$.
Then the distance contracting region, $\Omega$, 
is given  by the condition: $3/4 < x + y < 5/4 $,
where $(x,y)$ $\in$ ${\cal{S}}$.
That is, $\forall$  $(x,y) \in \Omega$: \,\, $d_{l}(x,y)$ $<$ $|x-y|$.  
If we require that the reduction of the distance is smaller than
$\varepsilon$ {\em i.e.} $d_{l}(x,y) < \varepsilon\,|x-y|$ 
then the states 
$(x,y)$  must be restricted to a strip $\Omega_{\varepsilon}$ $\subset $ 
$\Omega$ given by the inequality  $1- \varepsilon/4 < 
x + y < 1+  \varepsilon/4 $. The area  
$\tilde{\Omega}_{\varepsilon}$ $\subset$ $\Omega_{\varepsilon}$, where 
shrinking of the distance is being  the strongest,
 is found around $(x,y) = (1/2,1/2)$,  where 
$d_{l}(1/2 \pm \varepsilon /2, 1/2 \pm \varepsilon' /2 )$ $=$ 
$|\varepsilon ^{2}  - (\varepsilon')^{2}|$.

To proceed further we assume  that the random orbits are described by 
the joint   probability distribution function $\rho_{W}(x,y)$ ($(x,y) \in
\cal{S}$) and  apply the noise to all pairs of numbers 
$(x,y)$. Now   the probability   
$P_{W}(\varepsilon)$ that this procedure yields two 
numbers such that   after applying the map, they are moved  
into the contracting area ${ \Omega} _{\varepsilon}$ is given by the
formula  
\begin{equation}\label{probability}
P_{W}(\varepsilon) =  \int_{0}^{1} \int_{0}^{1} dx dy \rho_{W}(x,y) 
           \frac{ |{\cal{I}}(x,y;W) \cap { \Omega} _{\varepsilon} |}
                  {|{\cal{I}}(x,y;W) \cap \cal{S}| }
\end{equation}
where  $ {\cal{I}}(x,y;W) $ is the interval from ($x-W,y-W$) to
($x+W,y+W$) of length  $ |{\cal{I}}(x,y;W)| $ $=$ $2\sqrt{2}W$.
The key quantity entering the formula (\ref{probability}) 
is  $\rho_{W}(x,y)$.  
Clearly, $\rho_{W}(x,y)$ should  be  peaked 
about the diagonal $x=y$ with {\em strong} (absolute) maximum at
$(x,y) = (1,1)$.  This observation follows from the fact
that the  logistic map exhibits singular density of states 
near $1$ and $0$
and that the points around $0$ are images of those around $1$. As after 
each iteration the noise is added (with the constraint that the pair of
numbers stays within the basin of attraction of the map) 
the points around $0$ are obtained less frequently.  
Thus, to get maximal probability $P(\varepsilon)$ the 
integration in (\ref{probability}) should include the region 
around $(1,1)$  which implies that  $W \gtrsim 0.5$.  
On the other hand the second term under integral, 
which gives the probability of finding the noise-shifted 
point in $\Omega_{\varepsilon}$, is for $W \gtrsim 0.5$ a 
decreasing function of $W$ 
and saturates for $W>1$.  Combining these two facts we  conclude
that $P_{W}(\varepsilon)$  should be, for $W \lesssim 1$, a 
decreasing function of $W$,  approaching
saturation for $W>1$. The saturation could be proceeded by a maximum,
laying between  $0.5 \lesssim W \lesssim 1 $.
In this case one gets minimum of $T$ for $0.5 \lesssim W \lesssim 1 $.
Finally, $\lim_{\varepsilon \to 0}P_{W}(\varepsilon) 
\rightarrow 0$ implies that  $T \to \infty$ for $N \to \infty$.

The upper limit  of  $T$  for $W \gtrsim 0.5$ also could  be found from
the formula (\ref{probability}).  As the distribution $\rho_{W}(x,y)$
is strongly peaked around $(1,1)$ the function $P_{W}(\varepsilon)$
actually gives a probability of finding a point in the 
area $\tilde{\Omega}_{\varepsilon}$.
From  (\ref{probability}) it follows that the frequency of getting 
a point  in $\tilde{\Omega}_{\varepsilon}$ is 
proportional to $\varepsilon$ which
gives $T \sim 1/\varepsilon$. 
For $\varepsilon = 10^{-N/2}$  the  image 
of any point from $\tilde{\Omega}_{\varepsilon}$ yields two 
numbers that are  
$10^{-N}$ apart indicating that the two trajectories just  collapsed.
 Hence   $T\sim 10^{N/2}$,  which is what we find for $W \approx 1$.

These observations have important  practical consequences.
They suggest that  the coalescence time to a fixed  precision  could 
be controlled by the noise. For the map (\ref{logistic}) 
this is achieved by accepting only these $\eta$'s in (\ref{logistic}) which 
give points $(x',y')$  restricted to $\Omega_{\varepsilon}$. 
Allowed values of $\varepsilon$ as functions of assumed  $N$ and $T$ 
and for $W \ge 0.5$ could again be inferred from Eq.(\ref{probability}). 

In conclusion,  we showed that in order to understand  
the recently observed  phenomenon of coalescence of the trajectories 
\cite{pecora,maritan}
the precision of the calculations must be taken into account.
For all systems studied in the literature so far, 
the average coalescence time $T$ is either
linear or exponential function of precision  of the calculations.
Thus, statistically, coalescence  never occurs when precision is infinite.
Clearly the arguments as given are of general validity 
and the logistic map or the tent map could serve as  examples. 

One may wonder as whether predicted linear (or exponential) 
behaviour of $T$ with $N$ could be correlated with the sign of the 
maximal Lyapunov exponent.  If the identical chaotic systems are subject 
to an external noise, which is generated at each time step independently  
of the previous values and of the  states of the systems, than the analysis
proposed by Pikovsky applies \cite{pikovsky0}. In this case
the largest Lyapunov exponent can be calculated from a single system. 
If this exponent is negative than, at least,  for the systems studied in 
\cite{pikovsky0} this would imply the linear dependence of $T$ with $N$
({\em{}i.e.} synchronization). However, such  linear dependence 
 does not necessary mean that the  maximal Lyapunov 
exponent must be negative. For example, in the case of  the binary 
tent map, where $T$ varies linearly with $N$ ($ \forall g$), 
the sign of the Lyapunov exponent (for $N \to \infty$) could be
arbitrary. Also we would like  to add that  
the technique \cite{pikovsky0} does not apply 
when the noise depends on the states of the systems, 
Eqs.(\ref{logistic},\ref{shift}). 
In this case the difference between ensembles with positive and negative
Lyapunov exponents is highly nontrival \cite{paladin}. This issue is 
currently under investigations and will be reported elsewhere.
 
Finally,  let us note that  $T$ is governed by  
the properties of the invariant density of the chaotic map, 
the structure of the distance  stretching 
area(s), and the way in which the noise is introduced. 
This is demonstrated for the random logistic map (1), where $T$
varies exponentially with $N$. The law we found is  different from 
a simple  $ (10^{-N})^{-2}$  dependence 
suggested in \cite{pikovsky} ($\equiv \varepsilon ^{-D}$, where $D \equiv
2 $ is the topological dimension of the space). 

%
%
\acknowledgments
This work was supported  by the Conselho Nacional de Desenvolvimento
Cient\'{\i}fico e Tecnol\'ogico (CNPq) and 
by the Financiadora de Estudos e Projetos (FINEP) in  Brasil.
%
%

%
%
\begin{figure}
\caption[]{Logarithm of  average relaxation time 
versus accuracy for $W = 0.7$. 
The straight line represents best least-square fit obeying  $ 3 \le 
N \le  16$. Triangles represent absolute accuracy calculations while
circles relative ones. For each accuracy average is 
performed  over 100 samples. Insets show the histograms of the coalescence
times for $N=8$ and   for two sample sizes equal 100 and 10000,
respectively.}
\end{figure}
%
\begin{figure}
\caption[]{Logarithm of the average relaxation time versus 
the strength $W$ 
of the noise. Accuracies considered are  $N = 3$ (triangles and circles), 
4 (losangles), 5 (pentagons)  and 6 (hexagons).  
Averages are performed  over 100 samples except for  $N = 3$ 
(circles)  which corresponds to 1000 samples. 
The lines are introduced to  guide eyes.}
\end{figure}
%
\begin{figure}
\caption[]{Coefficients  $ A(W)$ (thin line)  and $B(W)$ (thick line)
of T(N) for the model (1).  The lines are
guide-to-eyes.}
\end{figure}
 
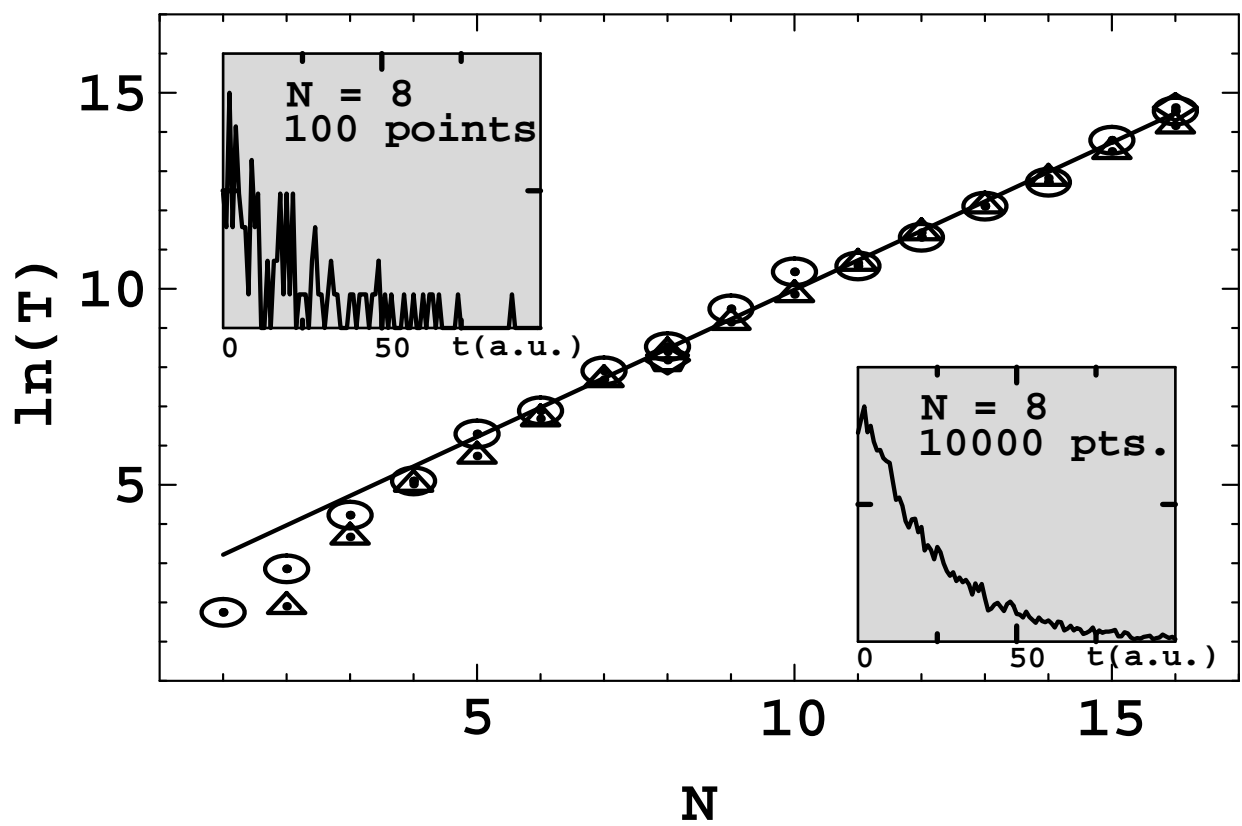
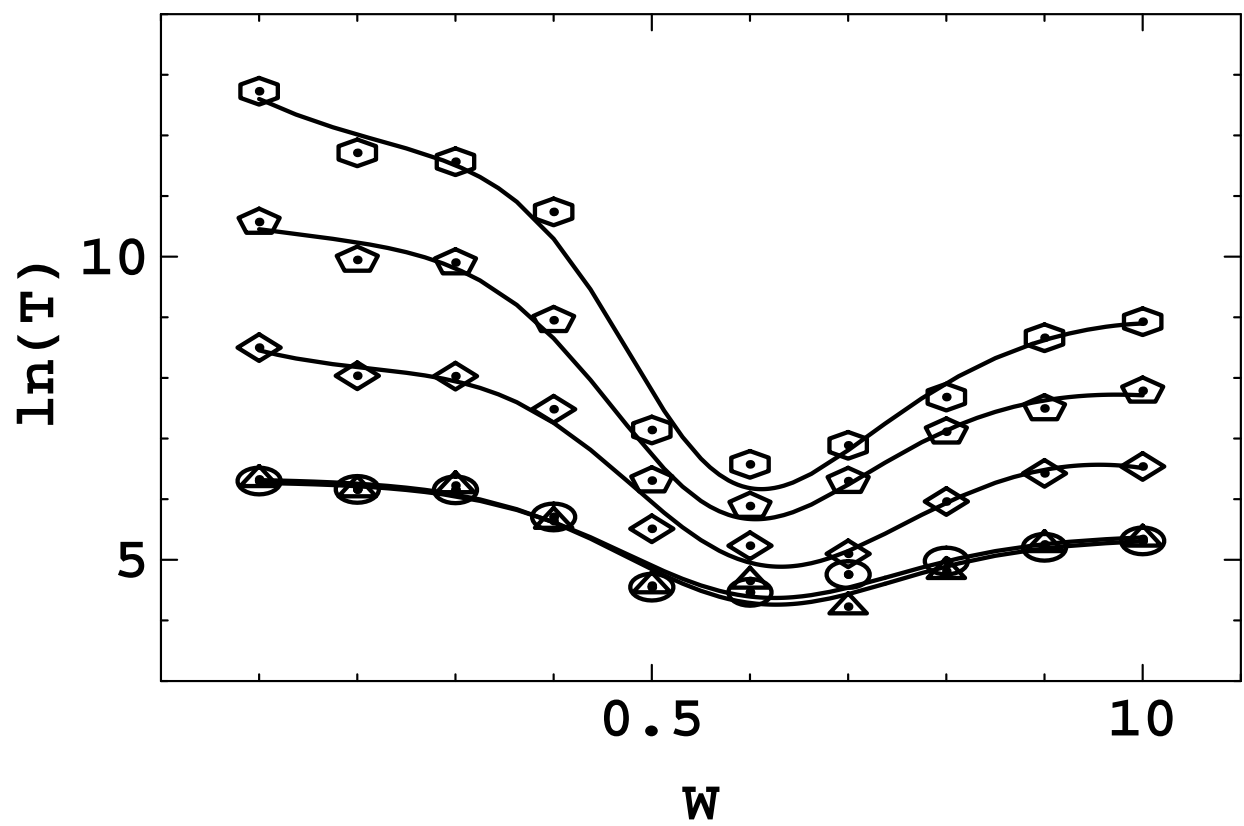
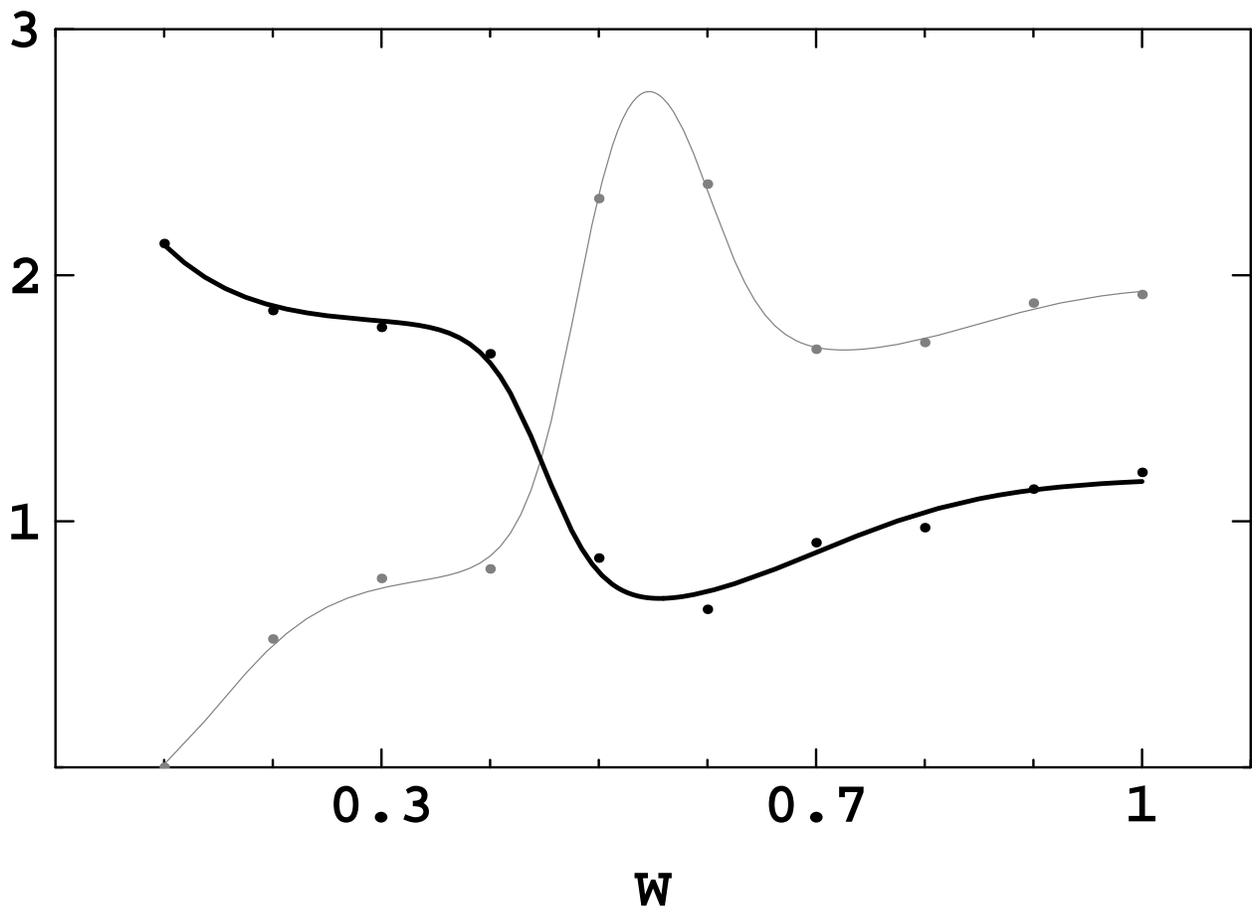
\end{document}